\documentclass[twocolumn,showpacs,preprintnumbers,nofootinbib,amsmath,amssymb]{revtex4}
\usepackage{amsmath,amssymb,amsfonts}
\usepackage{graphics}
\usepackage{epsfig}

\newcommand\al{\alpha}

\newcommand\de{\delta}

\newcommand\De{\Delta}
\newcommand\Ga{\Gamma}

\newcommand\Om{\Omega}
\newcommand\half{{\frac{1}{2}}}

\newcommand\ce{{\cal E}}

\newcommand\ep{\epsilon}
\newcommand\vep{\varepsilon}
\newcommand\MD{\mathfrak{D}}

\newcommand\tal{{\tilde{\alpha}}}
\newcommand\vak{\varkappa}

\begin{document}
\title{Discrete Wave-Packet Representation in Nuclear Matter Calculations}
\author{H. M{\"u}ther}
\email{herbert.muether@uni-tuebingen.de}
 \affiliation{Institute for Theoretical Physics, University of T\"ubingen, Auf der
Morgenstelle 14, D-72076 T\"ubingen, Germany}
\author{O.A. Rubtsova}
\email{rubtsova@nucl-th.sinp.msu.ru}
\author{V.I. Kukulin}
\email{kukulin@nucl-th.sinp.msu.ru}
\author{V.N. Pomerantsev}
\email{pomeran@nucl-th.sinp.msu.ru}
 \affiliation{Skobeltsyn Institute of
 Nuclear Physics, Moscow State University, Leninskie gory 1(2), Moscow,
119991, Russia}

\begin{abstract}
The Lippmann--Schwinger equation for the nucleon-nucleon $t$-matrix
as well as the corresponding Bethe--Goldstone equation to determine
the Brueckner reaction matrix in nuclear matter  are reformulated in
terms of the resolvents for the total two-nucleon Hamiltonians
defined in free space and in medium correspondingly. This allows to
find solutions at many energies simultaneously by using the
respective Hamiltonian matrix diagonalization in the stationary wave
packet basis. Among other important advantages, this approach
simplifies greatly
 the whole computation procedures
both for coupled-channel $t$-matrix and the Brueckner reaction
matrix. Therefore this principally novel scheme is expected to be
especially useful for self-consistent nuclear matter calculations
because it allows to  accelerate in a high degree single-particle
potential iterations. Furthermore the method provides direct access
to the properties of possible two-nucleon bound states in the
nuclear medium.
 The  comparison between reaction matrices
found via the numerical solution of the  Bethe--Goldstone integral
equation and the straightforward Hamiltonian diagonalization shows a
high accuracy of the method suggested. The proposed fully discrete
approach opens a new way to an accurate treatment of two- and
three-particle correlations in nuclear matter on the basis of
three-particle Bethe--Faddeev equation by an effective Hamiltonian
diagonalization procedure.
\end{abstract}
 \maketitle
\section{Introduction}
The conventional treatment of quantum  problems with continuous
spectrum (e.g. few-body scattering problems) uses either  the
differential Schroedinger-type equation with the appropriate
boundary conditions
 or employs an integral equation of the
Lippmann--Schwinger- or Faddeev-type \cite{newton}. A rather similar
formalism is used in the nuclear matter calculations within the
Brueckner--Bethe--Goldstone approach
\cite{Tabakin,Muether99,Baldo,book,Muether_02,Sartor} for finding
the Brueckner
 reaction matrix to treat strong two- or few-particle correlations in
many-nucleon system.

However, there are alternative approaches in quantum scattering
which allow to evaluate scattering observables as well as the fully
off-shell  transition operator by using some spectral properties of
the Hamiltonian. In this alternative way, one employs a
finite-dimensional approximation for the spectral expansion of the
total resolvent within the continuum discretization technique
\cite{Annals} or uses another $L_2$-type approach
\cite{Heller,reinhardt,papp,LIT}.
 By using this approximation,  the
initial  Lippmann--Schwinger  equation for the scattering $t$-matrix
is rewritten in a form which includes the  total  resolvent
expressed as a finite sum over the total Hamiltonian bound- and
pseudostates. Thus, a single Hamiltonian matrix diagonalization
makes it possible to determine scattering observables for all
necessary energies in a very broad interval. This discrete way for
solving a single-channel scattering problem becomes even  more efficient
in a coupled-channel case because it results in getting the multi-channel
  off-shell $t$ matrix  at many energies simultaneously.

Keeping in mind that the Bethe--Goldstone equation (BGE) for the
reaction matrix in nuclear matter differs from the
Lippmann--Schwinger equation (LSE) by the presence of the
Pauli-projection operator in its kernel, one can try to replace the
solution of the BGE also with a straightforward diagonalization
procedure of the respective Hamiltonian matrix calculated on the
appropriate basis.

The specific  feature of the nuclear matter calculations  is that
one has to solve the respective integral BGE
 many times to obtain the reaction matrix for
the various values of relative and center-of-mass momenta  and
energies needed to evaluate e.g. the single-particle potential in a
derivation of  the equation of state (EOS) \cite{Muether99,Tabakin}.
We keep in mind that  one employs usually an iterative procedure to
calculate the single-particle potential in a self-consistent way.
Thus, to find the EOS  one has to carry out quite a few iteration
steps to attain a self-consistent solution for solving the
Bethe--Goldstone integral equation at many energies
\cite{Muether99}. These self-consistent iterations are
time-consuming but still feasible on modern computer systems. The
numerical efforts, however, increase considerably if one tries to
account for an accurate treatment  of { three-body correlations} in
nuclear matter using two- and three-body forces on the basis of the
Bethe--Faddeev equations \cite{Sartor}. Therefore it is highly
desirable to develop efficient tools for the solution of these
equations.

Thus, in the present paper we employ the wave-packet continuum
discretization technique \cite{Annals} to carry out single- and
coupled-channel scattering calculations with realistic $NN$
interaction in vacuum as well as in medium by using
finite-dimensional approximation for the total resolvent. This
technique leads to an efficient way of solving the two-nucleon (NN)
problem with continuous single-particle spectra. In addition, this
method will also give explicit access to the properties of the bound
and quasi-bound states of two nucleons in vacuum as well as in
infinite matter (see e.g. the discussion of possible di-neutron
states in neutron matter \cite{dingri2016} and references cited
there).

The structure of the paper is the following: In Section II, we
explain in detail
 our approach to finding fully off-shell $t$-matrix at
many energies simultaneously
 via a
direct diagonalization of the total Hamiltonian matrix by using the
basis of stationary wave packets. Further, in Section III, this
approach is generalized to the solution for the Bethe--Goldstone
equation in nuclear matter. This alternative way turned
out to be simpler and faster as compared to the conventional
approach as is demonstrated in  Section IV which contains some
illustrative examples.  This section also includes a discussion of
bound two-nucleon states in the medium of nuclear matter. The
summary is given in Section V.  For the reader's convenience we have
added  the Appendix with some details of the discrete
coupled-channel formalism.

\section{Single- and coupled-channel scattering problem in a discrete representation}
\subsection{$t$-matrix for a two-body scattering problem}
Let us start with a reminding  some well-known formulas from the
quantum scattering theory \cite{newton}. The integral
Lippmann-Schwinger equation for the transition operator in momentum
representation ($t$-matrix)
    corresponding to a scattering of two particles
with momenta ${\bf k}_1$ and ${\bf k}_2$ has the following form:
\begin{eqnarray}
 t({\bf k},{\bf k'};E)=V({\bf k},{\bf
k}')+\nonumber\\
+\int\!\frac{{d^3k^{''}}V({\bf k},{\bf k}^{''})t({\bf k}'',{\bf
k}';E)}{E+{\rm i}0-\frac{(k'')^2}{2\mu}},\label{LSE}
\end{eqnarray}
where ${\bf k}={\bf k}_2-{\bf k}_1$ (and also ${\bf k'}$, ${\bf
k}''$) is the relative momentum of the particles,  $V$ is an
interaction potential,  $\mu$ is the reduced mass and the center of
mass momentum dependence is assumed to be  separated out.

In operator form, eq.~(\ref{LSE}) is written
\begin{equation}
\label{tm_int} \hat{t}(E)=\hat{V}+\hat{V}\hat{g}_0(E)\hat{t}(E),
\end{equation}
where $\hat{g}_0(E)=[E+{\rm i}0-\hat{h}_0]^{-1}$ is the resolvent of
the free Hamiltonian $\hat{h}_0$ (the kinetic energy operator) or
the Greens function for the non-interacting particles.

 Alternatively one
can introduce the  resolvent $\hat{g}(E)=[E+{\rm i}0-\hat{h}]^{-1}$
of the total Hamiltonian $\hat{h}=\hat{h}_0+\hat{V}$ including the
interaction. This total resolvent is related to the free resolvent
$\hat{g}_0(E)$ by the well-known identity:
\begin{equation}
\label{res_id}
\hat{g}(E)=\hat{g}_0(E)+\hat{g}_0(E)\hat{V}\hat{g}(E).
\end{equation}
If one knows the total resolvent $\hat{g}(E)$, the transition operator
(\ref{tm_int})  can be evaluated straightforwardly from the
relation:
\begin{equation}
\label{vgv} \hat{t}(E)=\hat{V}+\hat{V}\hat{g}(E)\hat{V}.
\end{equation}
At first glance the evaluation of the total resolvent from the
integral eq.~(\ref{res_id}) seems to be a more complicated task than
the calculation of the half-shell $t$-matrix through the integral
equation (\ref{tm_int}) at a single fixed energy. However, there are
some $L_2$ techniques \cite{Heller,reinhardt,papp,LIT,Annals} which
allow to find a finite-dimensional approximation  for $\hat{g}(E)$
and calculate the $t$-matrix elements directly using the relation
(\ref{vgv}).

These finite-dimensional approximations are usually based on a
spectral expansion of the total resolvent using a complete set of
states for the total Hamiltonian $\hat h$, i.e. using its bound
$\{|\psi_n^\al\rangle\}$ and continuum states
$\{|\psi^\al(E)\rangle\}$:
\begin{eqnarray}
 \hat{g}(E)=\sum_{\al}\sum_n\frac{|\psi_n^\al\rangle\langle
\psi_n^\al|}{E-E_n}+\nonumber\\
+\sum_{\al}\int_{0}^\infty dE'\frac{|\psi^\al(E')\rangle \langle
\psi^\al(E')|}{E+{\rm i}0-E'},\label{g_sp}
\end{eqnarray}
where $\al$ are appropriate quantum numbers referring to operators
which commute with the Hamiltonian. Approximating the spectral
expansion (\ref{g_sp})  by a finite sum over the Hamiltonian $\hat h$
pseudostates $\{|{z}_k^\al\rangle\}$ found in some appropriate $L_2$
basis \cite{Heller,Annals}, one gets the following
finite-dimensional form for the total resolvent $g(E)$:
\begin{eqnarray}
 \hat{g}(E)\approx
\sum_{\al}\sum_n\frac{|\psi_n^\al\rangle\langle
\psi_n^\al|}{E-E_n}+\nonumber\\+\sum_{\al,k}|{z}_k^\al\rangle
g_k^\al(E) \langle {z}_k^\al|.\label{g_pseudo}
\end{eqnarray}
Here $g_k^\al(E)$ are   complex functions  which depend on energy
$E$ and pseudostate energies $\{E_k^\al\}$ only. So this expansion
allows to find the resolvent $\hat{g}(E)$ and, furthermore, the
off-shell $t$-matrix at any energy $E$ by using only {\em a single
diagonalization of the total Hamiltonian matrix} for each channel
$\al$ (only functions $g_k^\al(E)$ should be recalculated which is
straightforward).

It should be mentioned that this ``spectral''\ scheme  is very
convenient because it allows to use any complete system of
eigenstates of the total Hamiltonian, such as standing wave
scattering functions
 \cite{newton} which are real valued.
In  other words, the spectral expansion does not require an accurate
treatment of boundary conditions for states in  the numerator of
(\ref{g_sp}) (that is why the pseudostates can successfully be used
here). Below, we use for this purpose the method
of the wave-packet continuum discretization (WPCD) developed by
present authors in previous years \cite{Annals}. This general
approach has been demonstrated \cite{Annals,KPRF} to be fully
applicable  for a coupled-channel interaction case as well (see also
the Appendix to the present paper).

One of the main purposes   of this paper is a careful testing of the
diagonalization procedure for construction of the coupled-channel
$t$-matrix for some realistic cases and also the generalization of
such a diagonalization approach to the solution of the
Bethe--Goldstone type equation widely used in nuclear matter
calculations. So that, below we discuss the case of $NN$ scattering
with the realistic interaction including  strong tensor components
which results in coupled-channel equations.
 However, the  scheme discussed is applicable also
for other types of coupled-channel problems, e.g. arising in a
coupled-channel reduction for some few-body scattering problems
\cite{validity}.

\subsection{Solution of a coupled-channel Lippmann--Schwinger equation in
 a wave-packet representation}
In this development, we apply a partial-wave expansion for
antisymmetrized wave functions and operators over the plane wave
states  $| k\rangle$ with definite orbital momentum $l$, spin $s$
and total angular momentum $j$ for the interacting nucleons (the
isospin value $\tau$ is determined by the $l$ and $s$ values, so we
omit it here to simplify notations).

 Then, after the partial-wave  projection, the
integral LSE (\ref{LSE}) takes the following form for given values
of $j$ and $s$ (those are conserved for the $NN$ Hamiltonian):
\begin{eqnarray}
t^{js}_{ll'}(k,k',E)  =  V^{js}_{ll'}(k,k')+\label{LSE_l}\\ +
\sum_{l''=|j-s|}^{j+s}\int_0^{\infty}
\frac{dk''V^{js}_{ll''}(k,k'')t^{js}_{l''l'}(k'',k',E)}{E-\frac{(k'')^2}{2\mu}+i0}
\nonumber.
\end{eqnarray}
Apart of the total angular momentum and spin, the parity is also
conserved for the $NN$ potential. Thus,  the eq.~(\ref{LSE_l})
 is  either a set of two coupled integral equations
for the spin-triplet case $s=1$ (for $l,l'=|j-1|,j+1$) or an
uncoupled equation for the spin-singlet ($s=0$) and the spin-triplet
($s=1$) channels with $l=j$.

To find the single- and coupled-channel $t$-matrices, we introduce
at first a partition of the continuum for the free Hamiltonian
$\hat{h}_0$ onto non-overlapping energy intervals (bins)
$\{\MD_i\equiv[\ce_{i-1},\ce_i]\}_{i=1}^N$ (or the corresponding
momentum bins $[k_{i-1},k_i]$)\footnote{We denote energy and
momentum intervals with the same notation $\MD_i$.} and then
construct
 the stationary wave-packet (WP) basis functions
$|x_i^l;\al\rangle$  as integrals of free plane
 waves $|k\rangle$ over these bins with inclusion of the necessary spin-angular parts
 $|\al\rangle\equiv|l,s:j\rangle$
 \cite{Annals}:
\begin{equation}
\label{xi}
|x_{i}^l,\al\rangle=\frac1{\sqrt{d_i}}\int_{k_{i-1}}^{k_i}dk|k,\alpha\rangle,\quad
d_i=k_i-k_{i-1}.
\end{equation}
 It is easy to see that the WP states
(\ref{xi}) form an orthonormal  basis:
\begin{equation}
\langle x_i^l,\alpha|x_{i'}^{l'},\alpha'\rangle =
\delta_{ii'}\delta_{\alpha\alpha'}.
\end{equation}
In such a basis, a simple finite-dimensional representation for the
free resolvent $\hat{g}_0(E)$ takes the form \cite{Annals}:
\begin{equation}
\hat{g}_0(E)=\sum_{\al}\sum_{i=1}^N |x_{i}^l,\al\rangle
g_i(E)\langle x_i^l,\al|,
\end{equation}
where the complex functions $g_i(E)$ can be found from the formula
(\ref{gie}) of the Appendix and depend on the discretization parameters $\ce_i$ and
total energy $E$ only \cite{Annals}.

Projecting out the eq.~(\ref{LSE_l}) onto such a discrete basis, one gets
the matrix equation for the coupled-channel
  $t$-matrix (and the corresponding eqs. for uncoupled cases):
\begin{eqnarray}
t^{js}_{il,i'l'}(E) &= &V^{js}_{il,i'l'}+
\label{lse_ml}\\&&+\sum_{l''=|j-s|}^{j+s}
\sum_{i''}V^{js}_{li,l''i''}g_{i''}(E)t^{js}_{l''i'',l'i'}(E),\nonumber
\end{eqnarray}
 where    $V^{js}_{il,i'l'}\equiv\langle
x_i^l,\al|\hat V|x_{i'}^{l'},\al'\rangle$ and
$t^{js}_{il,i'l'}(E)\equiv\langle
x_i^l,\al|{\hat t}(E)|x_{i'}^{l'},\al'\rangle$ are matrix elements of the
interaction and transition operator, respectively.

The off-shell $t$-matrix can be found from the solution of the
matrix eq.~(\ref{lse_ml}):
\begin{equation}
t^{js}_{ll'}(k,k';E)\approx\frac{
t^{js}_{il,i'l'}(E)}{\sqrt{D_iD_{i'}}},\quad
\begin{array}{c}
k\in\MD_i,\\
k'\in \MD_{i'},\\
\end{array}
\end{equation}
where $D_i=\ce_i-\ce_{i-1}$ are widths of the corresponding energy intervals .

 The respective $S$-matrix is derived from the diagonal element of
 the WP $t$-matrix:
\begin{equation}
S^{js}_{ll'}(E)\approx\de_{ll'}-2\pi{\rm
i}\frac{t^{js}_{il,il'}(E)}{D_i},\quad E\in\MD_i,
\end{equation}

However, such a procedure requires a separate matrix inversion at
every energy $E$ considered. So that, if one needs to evaluate the
coupled-channel $t$-matrix at many energies (e.g. for any Faddeev or
Faddeev--Yakubovsky calculation) the above direct way turns out
rather time-consuming. This problem is also important in case of
nuclear matter calculations where one has to calculate the reaction
matrix for various center-of-mass momenta and also to carry out the
numerous iterations to reach the self-consistency.

 Thus, below we will show how to  evaluate the off-shell $t$-matrix
without solving the scattering equations for each energy by using the
Hamiltonian matrix diagonalization in an appropriate model space.

\subsection{The resolvent of the total Hamiltonian}
For fixed values of $j$ and $s$, the total Hamiltonian can be
written as a  coupled-channel operator:
\begin{equation}
\label{h_ll}
\hat{h}^{js}_{ll'}=\hat{h}_{0l}^{js}\de_{ll'}+\hat{V}^{js}_{ll'},
\end{equation}
with a dimension $d$ equal to 1 or 2 for uncoupled and coupled cases
correspondingly.

The general idea of the wave-packet approach to find the total
resolvent of some Hamiltonian $\hat{h}$ is a discretization of its
continuum similarly to the free Hamiltonian case and a construction
of  {\em the scattering wave-packets} by the eq.~(\ref{zi}) from the exact scattering wave
functions. In this new WP representation, the resolvent of the
Hamiltonian $\hat{h}$ has a diagonal form with known eigenvalues
(see ref.~\cite{Annals} and the Appendix to the present paper). At
the next step, the scattering WPs are approximated by pseudostates
of the total Hamiltonian found in a free WP basis. Thus, by using
the same free WP basis and employing the total Hamiltonian matrix
diagonalization procedure, one can find finite-dimensional
approximations both for the free and total Hamiltonian resolvents $\hat{g}_0$ and 
$\hat{g}$ respectively.

Being fully straightforward for a single-channel total Hamiltonian,
this diagonalization approach requires some special consideration in
a coupled-channel case because the continuous spectra of the free and total
Hamiltonians are degenerated. So that, one has two different
scattering wave functions at each energy $E$ corresponding to
different initial states. The ordinary pseudostate technique does
not allow to distinguish these different channel states. However, we
have shown previously that special eigenchannel representation
formalism can be employed here \cite{Annals}.

Indeed, one can introduce
 two different types of scattering states for the
Hamiltonian (\ref{h_ll}) at given energy $E$: the scattering states
$|\psi^{l}(E)\rangle$ (which include also spin-angular parts)
corresponding to the incoming waves with a definite orbital momentum
$l$ and the scattering states $|\psi^{\varkappa}(E)\rangle$ defined
in the so-called eigenchannel representation (ER) (which corresponds
to a diagonal form of the total $S$-matrix \cite{Greiner,KPRF}),
where $\varkappa$ is an eigenchannel index. For example, in case of
the coupled-channel triplet $NN$ interaction, the scattering states
$|\psi^{\varkappa}(E)\rangle$ are linear combinations of the states
$|\psi^{l}(E)\rangle$, e.g. eigenchannel states for the coupled
channels $^3S_1-{}^3D_1$, $^3P_2-{}^3F_2$ etc.\footnote{In nuclear
physics, a conventional notation for these pair eigenstates is
adopted as the so-called $\alpha$ state and $\beta$ state. Here we
use the general index $\varkappa$.}

Further, by making use of the above eigenchannel representation in a coupled-channel case, one
gets the following spectral expansion for the resolvent of the total
Hamiltonian $\hat{g}(E)$:
\begin{equation}
\label{2g1} \hat{g}^{js}(E)=\frac{|z_b\rangle\langle
z_b|}{E-\epsilon_b}+\sum_{\varkappa=1}^d\int_0^\infty
\frac{|\psi^{\varkappa}(E)\rangle\langle\psi^{\varkappa}(E)|}{E+i0-E'}dE',
\end{equation}
which is {\em diagonal} with respect to the eigenchannel index
$\varkappa$. In the eq.~(\ref{2g1}),  $\epsilon_b$ is an energy of
the bound state $|z_b\rangle$ (the deuteron). In the $NN$ system, the bound-state term
takes place for the triplet with $j=1$ channel, only.

In order to use further the WP technique with the pseudostate
approximation, one has to prepare the discretization   partitions of
spectra in different coupled spin-angular channels  in such a way
that a coupled-channel free Hamiltonian matrix  would have  {\em a
degenerate discrete spectrum} \cite{Annals,KPRF}. After switching on
 the interaction, these multiple discrete energy levels are split and
form a set of paired levels (for two coupled channels \cite{KPRF}).
So that, one can group the resulted coupled-channel pseudostates
into two branches in the eigenchannel representation. We have shown
\cite{KPRF} that such pseudostates obtained in the free WP basis can
easily be related to scattering wave functions
$|\psi^{\varkappa}(E)\rangle$. More definitely, these pseudostates
can be considered as approximations for the {\em multichannel}
scattering wave packets constructed from  exact scattering wave
functions (defined in the ER)  of the coupled-channel total
Hamiltonian $\hat{h}_{ll'}$ (see the eq.~(\ref{z_vak}) in the
Appendix).

Thus, the diagonalization of the total Hamiltonian (\ref{h_ll})
matrix  in the two-channel
 free WP basis $\{|x_i^l;l,s:j\rangle\}_{l=|j-s|}^{j+s}$ results in a set
 of  pseudostates $|z_k^\vak,\tal\rangle$ with eigenenergies
 $E_k^\vak$,
 which are expanded in  free WP states:
\begin{equation}
\label{z_viax} |z_k^\vak,\tal\rangle = \sum_{i,l} C_{ki}^{\vak l}
|x_i^l,\al\rangle.
\end{equation}
Here  $\vak$ is an eigenchannel index and the total spin-angular
part has the form $\tal=\{\vak,s,j\}$.

With such a treatment of the multichannel pseudostates,  a
finite-dimensional representation for the total resolvent takes a
diagonal form \cite{Annals} both for a single- and a coupled-channel cases:
\begin{equation}
\label{Rtot} g^{js}(E)  \approx \frac{|z_b\rangle\langle
z_b|}{E-\epsilon_b}+\sum_{\vak=1}^d\sum_{k=1}^{N_{\vak}}
|{z}_k^\vak\rangle g_k^\vak(E)\langle {z}_k^\vak|.
\end{equation}
Moreover, the explicit formula for $g_k^\vak(E)$ is the same  for
both cases (see the Appendix).

Finally, the representation of the total resolvent as a finite spectral
sum, as in the eq.~(\ref{Rtot}), allows a direct evaluation of the
multienergy off-shell $t$-matrix.

After carrying out the  diagonalization procedure for the total
Hamiltonian matrix at a fixed coupled-channel partial wave ($j,s$),
the off-shell $t$-matrix is found from the explicit relation
(instead of solving the eq.~(\ref{lse_ml})):
\begin{eqnarray}
t^{js}_{il,i'l'}(E)\approx V^{js}_{il,i'l'}+\frac{V^{11}_{il,{\rm
b}}V^{11}_{i'l',{\rm b}}}{E-\ep_{\rm b}}\de_{j1}\de_{s1}+\nonumber\\
 +\sum_{\vak=1}^d
\sum_{k=1}^{N^\vak}{V^{js}_{il,k\vak}} g_{k}^\vak(E)
V^{js}_{i'l',k\vak},
\end{eqnarray}
where $d=2$ or 1  for coupled or uncoupled  channel, respectively.
 Here specific
 matrix elements of the interaction operator have been introduced:
$V^{js}_{il,{\rm b}}\equiv \langle x_i^l,\al|\hat V|z_{\rm
b}\rangle$ and $V^{js}_{il,k\vak}\equiv \langle x_i^l,\al|\hat
V|z_k^\vak,\tal\rangle$.
 They are calculated from the  matrix elements of the interaction in the initial free
WP basis by using the eq.~(\ref{z_viax}), e.g.:
 \begin{equation}
V^{js}_{il,k\vak}=\sum_{l',i'} C^{\vak l'}_{ki'}V^{js}_{il,i'l'}.
 \end{equation}

\begin{figure}
 \epsfig{file=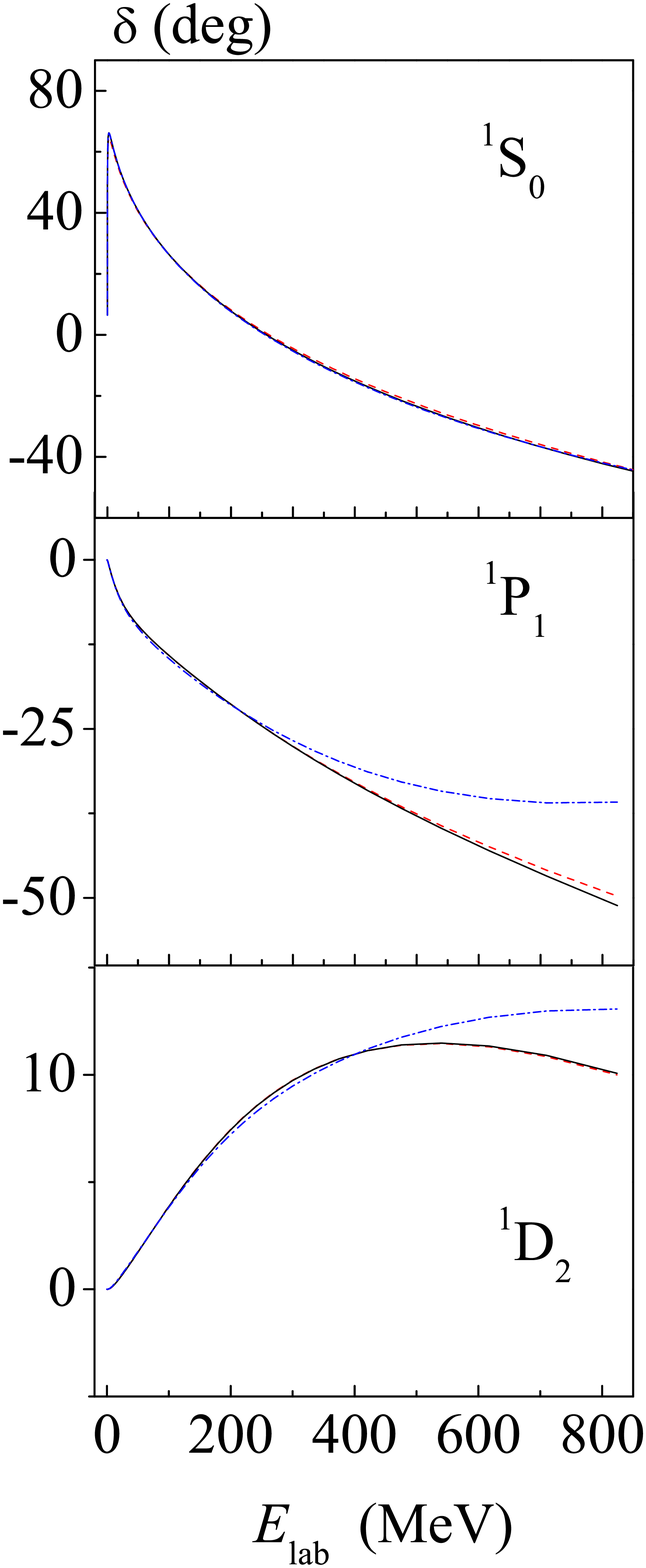,width=0.53\columnwidth}
\hspace{-1cm} \epsfig{file=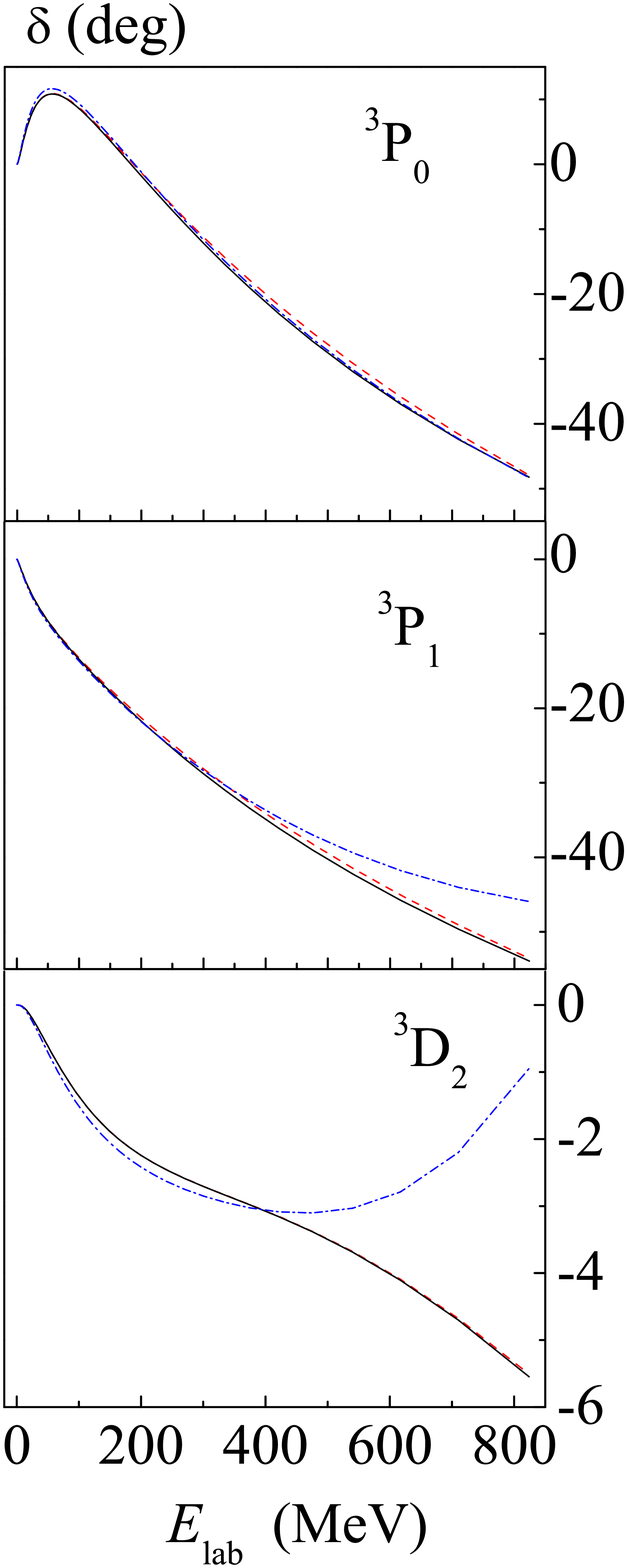,width=0.53\columnwidth}
\caption{\label{fig_singlet} (Color online) The partial phase shift
$\delta$ for uncoupled spin-singlet (left panel) and spin-triplet
(right panel) channels using the CD-Bonn $NN$ potential derived from
a total Hamiltonian diagonalization in the WP basis (solid curves)
and from the direct numerical solution of the one-channel
Lippmann--Schwinger equation (dashed curves). The dash-dotted curves
corresponds to partial phase shifts evaluated for the Nijmegen NN
potential.}
\end{figure}

As an example for the application of diagonalization procedure to
find the  NN scattering amplitudes we present in
Fig.~{\ref{fig_singlet}} the partial phase shifts for the uncoupled
spin-singlet  and spin-triplet channels using the $NN$ CD-Bonn
potential \cite{CDBonn}.  To check the accuracy of the approach, we
compare in this Figure the results of the direct solution for the
integral Lippmann--Schwinger equation in a matrix form
(\ref{lse_ml}) (which has to be solved at the various energies
independently) with results of a single diagonalization for the
respective $NN$  coupled-channel Hamiltonian in very broad energy
interval of laboratory energy $E_{\rm lab}$ from zero up to 800 MeV.
It can be seen that the results for the direct and diagonalization
solutions
 are almost indistinguishable from each other  in the whole
energy  region studied. We have also added to the
Fig.~{\ref{fig_singlet}}  the partial phase shifts evaluated for the
Nijmegen II NN potential \cite{Nijm}. The phase shifts derived from
the different potentials agree mainly in the energy region, in which
they both were fitted to the experimental data. The different
behavior of the phase shifts for energies above 350 MeV is
well-known.

\begin{figure}
 \epsfig{file=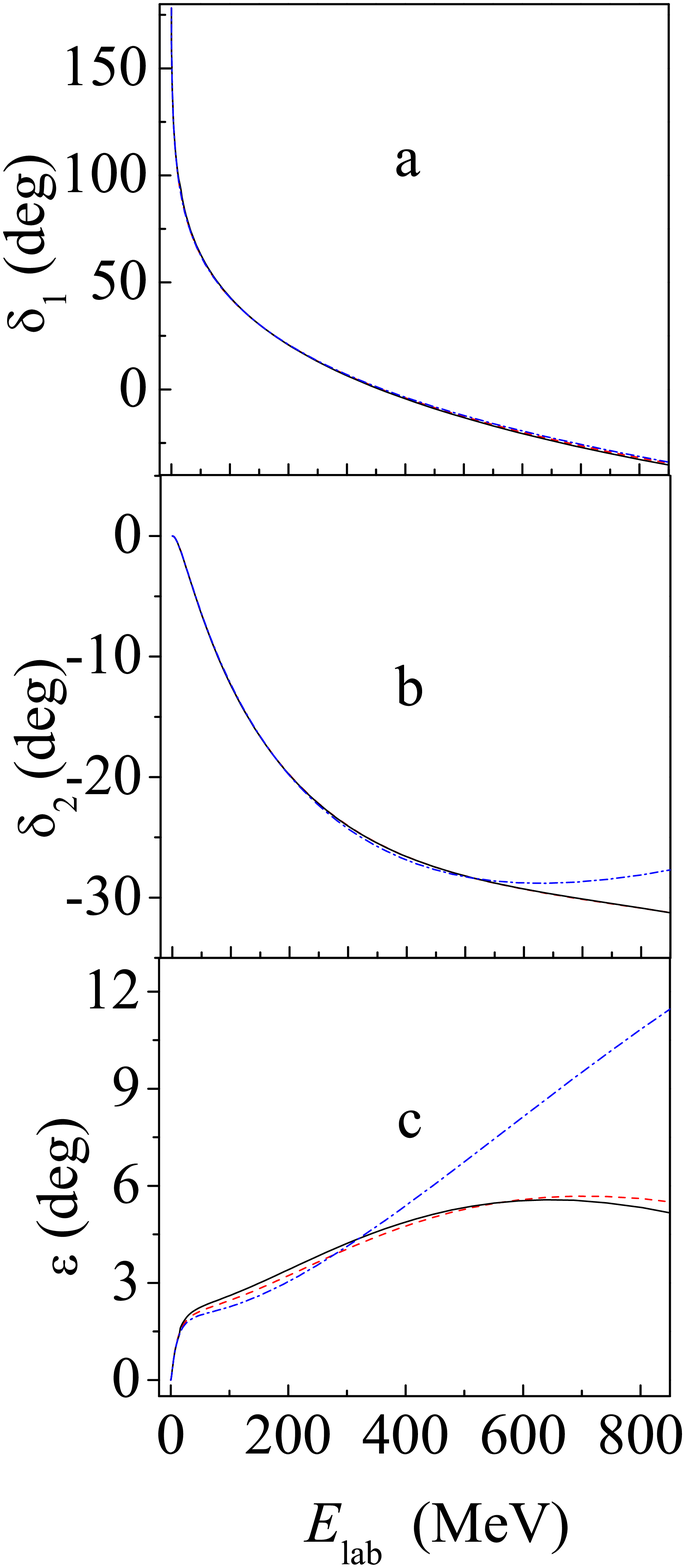,width=0.53\columnwidth}
\hspace{-0.8cm} \epsfig{file=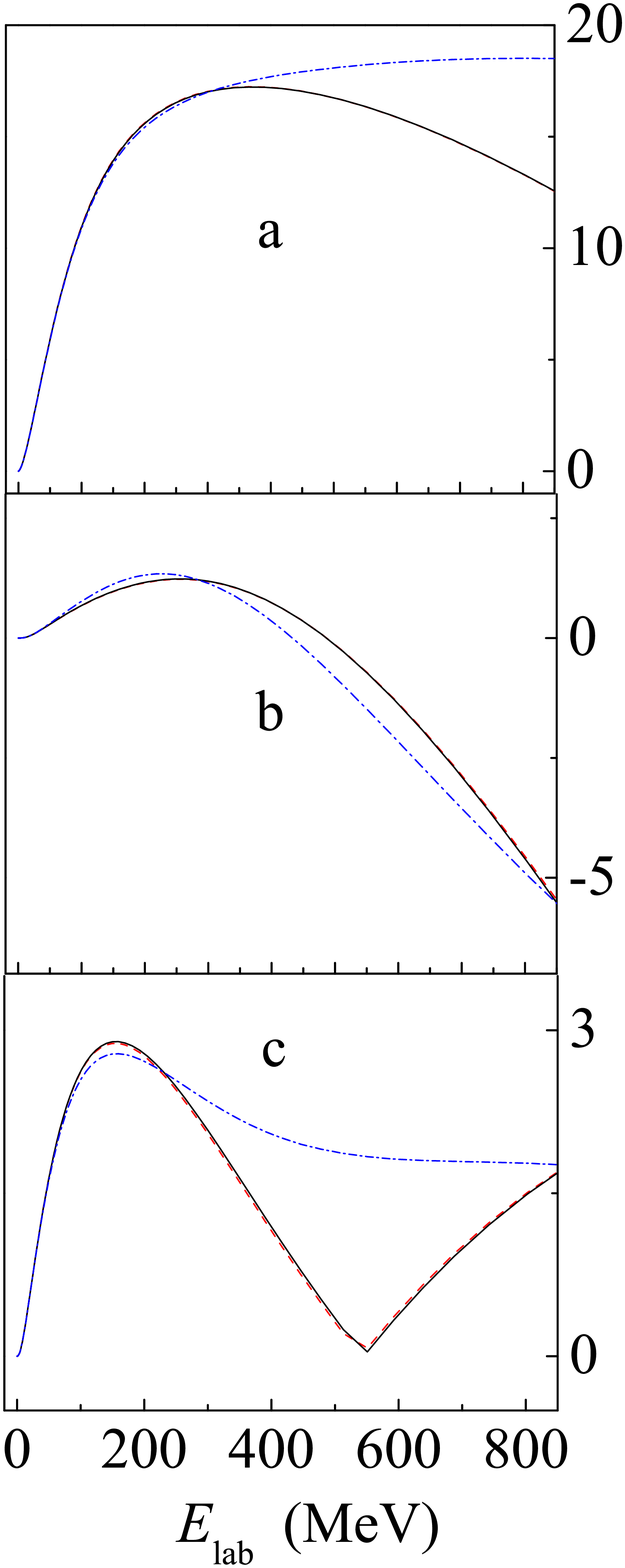,width=0.53\columnwidth}
\caption{\label{fig_triplet} (Color online) The partial phase shifts
$\delta_1$ (a), $\delta_2$ (b) and the mixing parameter $\vep$ (c)
for the coupled spin-triplet channels ${^3}S_1-{^3}D_1$ (left panel)
and ${^3}F_2-{^3}P_2$ (right panel) of the $NN$ scattering found for
the
 CD-Bonn and Nijmegen $NN$ potentials. The notations of curves are the same as in
 Fig.~\ref{fig_singlet}.}
\end{figure}

In Fig.~\ref{fig_triplet} the coupled-channel  phase shifts and
mixing angle $\varepsilon$ (in the Stapp parametrization) for the
spin triplet channels of $NN$ scattering with CD-Bonn potential are
displayed and compared to solutions of the Lippmann--Schwinger
equation in a matrix form eq.~(\ref{lse_ml}). The very good accuracy
which the diagonalization technique can reach in scattering
calculations is obvious also for the coupled-channel cases.

The clear advantage of the diagonalization technique as compared to
the conventional solution of the LSE equation is evident, in
particular, if one has to solve the LSE at many energies as it is in
the case e.g. for the calculation of integral kernels of
Faddeev--Yakubovsky three- and many-body integral equations. As an
example we refer to the three-body discrete Faddeev calculations
\cite{Annals}, where we have used  the finite-dimensional
approximation for the total resolvent~(\ref{Rtot}) derived from the
Hamiltonian diagonalization procedure.

\section{Solving the Bethe--Goldstone equation by a matrix diagonalization}

Let us consider a case of infinite symmetrical nuclear matter at
zero temperature, where  we study  the respective Bethe--Goldstone
integral equation for the Brueckner reaction matrix
\cite{Muether99,Tabakin}:
\begin{eqnarray}
T({\bf k},{\bf k}';K,W_0)=V({\bf k},{\bf k}')+\int{d^3k^{''}}V({\bf
k},{\bf k}^{''})\nonumber\\
\times\frac{Q({\bf k}'',K)T({\bf k}'',{\bf
k}';K,W'')}{W_0+i0-H_0({\bf k}'',{\bf K})}\label{BG},\quad
\end{eqnarray}
where ${\bf K}=\half ({\bf k}_1+{\bf k}_2)$ and ${\bf k}=\half ({\bf
k}_2-{\bf k}_1)$ are the  center-of-mass (c.m.) and relative
momenta, respectively, $W_0$ is the starting energy, $H_0({\bf k}'',
K)$ defines the energy of the intermediate state with relative
momentum ${\bf k}'' $ and $Q({\bf k}'',K)$ is the Pauli-projection
operator:
\begin{equation}
\label{qkp} Q({\bf k},K)=\theta(|{\bf k+K}|-k_F)\theta(|{\bf
k-K}|-k_F),
\end{equation}
which excludes particle states forbidden by the Pauli principle, so
that  $k_F$ is the Fermi momentum. Due to the appearance of the
Pauli projection operator in the kernel of the eq.~(\ref{BG}), the
c.m. momentum $\bf K$ is not separated out here (as it was for the
LSE) but it is still  conserved. The eq.~(\ref{BG}) is
traditionally solved in the momentum space formed by the relative
momentum $\bf k$ while the absolute  value of the c.m.
momentum $K$ in the eq.~(\ref{BG}) plays the role of an external
parameter, in addition to the starting energy $W_0$.

In most of the calculations,
 the so-called angle-average approximation for the Pauli operator $Q$
 is assumed \cite{Tabakin}:
\begin{equation}
Q(k,K)=\left\{
\begin{array}{lr}
0,& k\le k_0,\\
\min\left\{1,\frac{k^{2}+K^2-k_F^2}{2kK}\right\},& k> k_0,\\
\end{array}
\right.\label{angavq}
\end{equation}
where $ k_0=\sqrt{k_F^2-K^2}$. If furthermore one considers a
spectrum of particle-states defined by $H_0({\bf k}, K)$, which is
also independent on the angle between the corresponding relative
${\bf k}$ and c.m. momentum $\bf K$ the Bethe-Goldstone equation is
symmetric under rotation and can easily be solved in a partial wave
expansion (see below) with matrix elements of the resulting reaction
matrix diagonal in this partial wave basis with respect to $j$ and $s$. 
In the present study, we
will also take advantage of this angle-average approximation. Note,
however, that an extension of the formalism to an exact treatment of
$Q$ and an angle-dependent  energy term $H_0$ is straightforward.

It should be stressed, that the operator $\hat Q$  in this angle-average
approximation is not a  projection operator because it does not
satisfy the relation ${\hat Q}^2=\hat{Q}$, but it can be considered in the
  relative momentum space as an
operator $\hat{Q}(K)$  depending on the c.m. momentum value.

The energies in the denominator of the eq.~(\ref{BG})
 $W_0$ and $H_0(k,K)$  are usually defined through
the Brueckner Hartree Fock (BHF) single-particle (sp) energy
\begin{equation}
\varepsilon(k_1)=\frac{k_1^2}{2m}+{\rm Re}
\left(U(k_1,\omega=\varepsilon(k_1))\right),\label{epsofk}
\end{equation}
where the self-energy $U$ is defined in terms of the reaction matrix
itself:
\begin{equation}
\label{ukk} U(k_1,\omega)=\int_{k_2\leq k_F} d^3k_2 T({\bf k},{\bf
k};K,W_0=\omega+\varepsilon(k_2)).
\end{equation}
In the last equation, the integral over ${\bf k}_2$ should be
restricted to the hole states. i.e. single-particle states  inside the Fermi sphere \cite{Muether99}.

While the on-shell definition of the energy variable $\omega$ in eq.
(\ref{epsofk}) is well established for the single-particle energies
of the hole states by the Bethe--Brandow--Petschek theorem
\cite{BBP}, the corresponding choice for the spectrum of the
particle states with momenta larger than $k_F$, which are needed to
define $H_0$, has been a matter of a long discussion in the
literature. Two different choices have been discussed:
\begin{itemize}
\item The conventional choice representing the
single-particle energies for the particle states by the kinetic
energy only
$$ \varepsilon(k_1) = \frac{k_1^2}{2m}.$$
With this choice one obtains a gap at $k_1=k_F$ as the attractive
single-particle potential $U(k_1)$ defined in eq.~(\ref{ukk}) is
taken into account for the hole states ($k_1<k_F$) but is ignored
for the particle states.
 Therefore this choice has also been
denoted as the ``gap'' spectrum.
\item The ``continuous'' or gap-less choice, in which the definition of
the single-particle energy according to eq.~(\ref{ukk}) is also
applied to the particle states with $k_1>k_F$.
\end{itemize}

The discussion of the optimal choice for the single-particle
potential has essentially been settled with the observation of Baldo
et al. \cite{3-body} demonstrating that  the effect of three-nucleon
correlations is reduced considering the continuous choice. Therefore
this continuous choice has become the standard choice in BHF
calculations.

In any case one obtains a monotonic rise of the single-particle
energy $\varepsilon(k_1)$ as a function of the momentum  $k_1$ and
the matrix elements of $T$ are complex only for energies $\omega$
and corresponding starting energies $W_0$, which are needed for the
evaluation of the particle states. This implies that the BHF
single-particle potential (\ref{ukk}) is real for the hole states
with momenta $k_1<k_F$. For energies $\omega$ larger than the Fermi
energy $\varepsilon_F = \varepsilon(k_F)$ the single-particle
potential $U(k_1)$ yields an imaginary component and this complex
self-energy $U(k_1,\omega)$ has been used to evaluate the optical model
potential for nucleon-nucleus scattering \cite{mahaux}.

Note that the solution of the Bethe--Goldstone eq.~(\ref{BG}) and
the evaluation of the single-particle potential eq.~(\ref{ukk}) must
be done in a self-consistent way since the BGE
requires the knowledge of the single-particle spectrum and the
evaluation of the single-particle energies requires the knowledge of
the reaction matrix $T$, that is the solution of the
Bethe--Goldstone equation. The self-consistent solution can be
obtained in an iterative way recalculating in each iteration step
the single-particle spectrum until the resulting spectrum will agree
with the spectrum which is used in the Bethe--Goldstone equation.

In this iteration procedure the single-particle spectrum to be used
in eq.~(\ref{BG}) has often been parameterized in terms of an
effective mass $m^*$:
\begin{equation}
\varepsilon_{app}(k_1) = \frac{k_1^2}{2m^*} + U_0\label{efmas}.
\end{equation}
This parametrization simplifies the iteration procedure and ensures
that the energy denominators used in the Bethe-Goldstone
eq.~(\ref{BG}) do not depend on the angle between relative and c.m.
momentum of the interacting pair of nucleons. This is a very useful
feature together with the angle-average definition of the Pauli
operator (\ref{angavq}). It has been observed~\cite{Muether99},
however, that effective-mass parametrization is not very accurate
(see also discussion below).

When the iteration scheme discussed above is converged, the binding
energy per nucleon (the equation of state) can be found with the
resulting sp spectrum using the relation:
\begin{equation}
\frac{E}{A} = \frac{3}{k_F^3}\int_0^{k_F} k^2\,dk\,
\frac12\left[\frac{k^2}{2m} + \varepsilon(k)\right]\,.
\label{eq.EA}
\end{equation}

\subsection{The resolvent operator}
For a standard BHF calculation, it is sufficient to solve the
BGE by means of standard techniques for
solving integral equations as e.g. it has been introduced by Haftel
and Tabakin~\cite{Tabakin}. Calculations beyond  the standard BHF
calculations, like the evaluation of the self-energy beyond lowest
order hole-line expansion, or the solution of the three-body
Bethe--Faddeev equation or the evaluation of the fully off-shell
behaviour of the self-energy, all these require the determination of
the reaction matrix at various starting energies. Thus, the discrete
wave-packet representation should be useful for such multifold
calculations, also it allows to obtain a different view on the
solution of the Bethe--Goldstone equation. For this aim, we will take
a brief look at the propagator  for the Bethe--Goldstone equation or
the resolvent operator for the corresponding total Hamiltonian.

Let us introduce the free  Hamiltonian for  two non-interacting
nucleons in nuclear matter
 for a fixed value of the  c.m. momentum $K$:
\begin{equation}
\label{h_0} \hat H_0(K)=\int d^3 k |{\bf k}\rangle H_0({\bf k},K) \langle
{\bf k}|,
\end{equation}
where $|{\bf k}\rangle$ are plane wave state for the relative
momentum ${\bf k}$ and energy terms $H_0({\bf
k},K)=\varepsilon\left(|{\bf k}-{\bf K}|\right)+\varepsilon\left(|{\bf
k}+{\bf K}|\right)$ are defined by means of a given sp potential.

Then,  the eq.~(\ref{BG}) can be rewritten in the operator form
\begin{equation}
\label{BG_op} \hat{T}(K,W)=\hat{V}+\hat{V}\hat{Q}(K)\hat{G}_0(K,W)
\hat{T}(K,{W}),
\end{equation}
where  the free resolvent $\hat{G}_0(K,W)$ is defined as usually as
$\hat{G}_0(K,W)=[W+{\rm i}0-\hat{H}_0(K)]^{-1}$ for the free
Hamiltonian (\ref{h_0}). Below we will omit the explicit dependence
of operators on $K$ where  is possible in order to simplify
notations.

It can easily be proven   that  if one  introduces some operator
$\hat{G}(W)$ which satisfies the operator equation:
\begin{equation}
\label{g_Q}
\hat{G}(W)=\hat{Q}\hat{G}_0(W)+\hat{Q}\hat{G}_0(W)\hat{V}
\hat{G}(W),
\end{equation}
then the solution of the eq.~(\ref{BG_op}) (the reaction matrix)
can be found from the formal relation
\begin{equation}
\label{G_vgv} \hat{T}(W)=\hat{V}+\hat{V}\hat{G}(W)\hat{V},
\end{equation}
 similarly to the ordinary $t$-matrix case.

Thus, if one would find some convenient way for the evaluation of
the  operator $\hat{G}(W)$  then the reaction matrix could be
calculated straightforwardly from the eq.~(\ref{G_vgv}).

For this purpose,  let us introduce two orthogonal subspaces ${\cal
H}_\Ga$ and ${\cal H}_Q$ of the total momentum space ${\cal H}$ with
respect to an action of the operator $\hat{Q}(K)$. Here ${\cal
H}_\Ga$ is the null space of $\hat{Q}(K)$ (it includes the states
$|{\bf k}\rangle$ for which $\hat{Q}(K)|{\bf k}\rangle=0$), while
${\cal H}_Q$ is its orthogonal complement (i.e. it includes the
Pauli-allowed states). Below we will denote projections of the
operators onto these subspaces with additional subindex $\Ga$ or $Q$
respectively.

Because $\hat{Q}$ commutes with $\hat{G}_0(W)$, it becomes clear
from the eq.~(\ref{g_Q}) that $\hat{G}(W)$ commutes with $\hat{Q}$ too.
Thus, the eq.~(\ref{g_Q}) should be considered in the subspace
${\cal H}_Q$ only. Due to the definition,    the inverse operator
${\hat Q}^{-1}$ and also the operators  ${\hat Q}^{\half}$ and
${\hat Q}^{-\half}$ exist in this subspace.

Then the eq.~(\ref{g_Q}) can be rewritten in a symmetric form:
$$
{\hat Q}^{-\half}{\hat G}{\hat Q}^{-\half}={\hat G}_{0Q}+\hat{G}_{0Q}\
{\hat Q}^{\half}\hat{V}{\hat Q}^{\half}\ {\hat Q}^{-\half}{\hat
G}{\hat Q}^{-\half},
 $$
  which is similar to the
resolvent identity (\ref{res_id}). So that, it is straightforward to
derive the following explicit form for the $\hat{G}(K,W)$ operator:
\begin{equation}
\hat{G}(K,W)=\hat{Q}^\half\Big[W+{\rm i}0-\hat{H}_{0Q}(K)- {\hat
Q}^\half\hat{V}{\hat
Q}^{\half}\Big]_Q^{-1}\hat{Q}^{\half},\label{gkz}
\end{equation}
where the inverse operator is defined in ${\cal H}_Q$.  Further, one
can introduce the spectral expansion for  the operator
$\hat{G}(K,W)$ in term of eigenstates of the following effective
Hamiltonian $H_Q(K)$:
\begin{equation}
\label{HK}
\hat{H}_Q(K)=\hat{H}_{0Q}(K)+\hat{Q}^{\half}(K)\hat{V}\hat{Q}^{\half}(K)
\end{equation}
which is defined  in the ${\cal H}_Q$ (as well as $\hat{H}_{0Q}$ is
a part of the free Hamiltonian (\ref{h_0}) in ${\cal H}_Q$).

The operator (\ref{gkz}) is nothing else as  an analog of the total
resolvent operator $g(E)$ used in an ordinary scattering problem.
Finally, the reaction matrix can be found in this formalism using
the explicit formula:
\begin{equation}
 \hat{T}(K,W)=\hat{V}+\hat{V}\hat{Q}^{\half}
 \left[W+\!{\rm i}0\! - \hat{H}_Q(K)\right]_Q^{-1}\hat{Q}^{\half}\hat{V}.\label{GGK}
\end{equation}
which is very convenient because the energy and $K$-dependencies
are separated in it.

Thus, one can recognize in the final eq.~(\ref{GGK}) an analog of
the eq.~(\ref{vgv}) for the transition operator in free
space which relates  the $t$-matrix and the total resolvent $\hat g$. So,
in quite a similar manner, one can replace the multiple solutions of
the BGE~(\ref{BG}) at many values of relative momentum $k$ with a
single diagonalization of the Hamiltonian matrix $\hat{H}_Q(K)$ in
${\cal H}_Q$ subspace, no matter which particular form of the
Pauli-exclusion operator $\hat{Q}$ is used (i.e. the angle-averaged or
the exact form).
\subsection{Evaluation of the reaction matrix in the discrete WP representation}
Here we apply the  discrete formalism, developed in Section II, to
derive the reaction matrix using a partial wave decomposition in the
eq.~(\ref{GGK}). For our illustrative purpose, the angle averaged
approximation for $\hat{Q}(K)$ operator \cite{Tabakin}  is used. In
that case, its momentum eigenvalues do not depend on spin-angular
quantum numbers and have the form displayed in eq.~(\ref{angavq}).

 Thus, the ${\cal H}_Q$ subspace
includes  plane wave states with relative momentum $k>k_0$ (see
eq.(\ref{angavq})) and it is convenient to introduce the
discretization bins in such a way that the momentum $k_0$ should
coincide to the endpoint of some bin. Thus,  we have
$\{|x_i\rangle\}_{i=1}^{N_1}$ and $\{|x_i\rangle\}_{i=N_1+1}^N$ sets
as the bases for  subspaces ${\cal H}_\Ga$ and ${\cal H}_Q$
respectively and $k_{N_1}=\sqrt{k_F^2-K^2}$.

In  case of the angle-averaged projector, the matrix elements of the
operator $\hat{Q}(K)$ can be found as follows:
\begin{eqnarray}
\hat{Q}(K)=\sum_{\al}\sum_{i=N_1+1}^N|x_i,\al\rangle Q_i(K) \langle
x_i,\al| ,\nonumber\\
Q_i=\frac1{d_i}\int_{k_{i-1}}^{k_i} dk Q(k,K),\quad
d_i=k_i-k_{i-1}\label{q_i}
\end{eqnarray}
If furthermore we assume that also the eigenvalues of the
free Hamiltonian $\hat{H}_0$ are independent on the the
angle between relative and c.m. momenta, the matrix elements of the
effective Hamiltonian (\ref{HK}) take the form for a coupled-channel
$NN$ interaction:
\begin{equation}
[H]^{js}_{il,i'l'}=[H_0]_i\de_{ii'}\de_{ll'}+\sqrt{Q_i}V^{js}_{il,i'l'}\sqrt{Q_{i'}},
\quad\label{eqeffham}
\end{equation}
where $i,i'=N_1+1,\ldots,N,\quad l,l'=|j-s|,j+s$ (or $l=l'=j$ for uncoupled channels), and $[H_0]_i$ is
the matrix element of the free Hamiltonian $\hat H_0$.

The continuous spectrum of the effective Hamiltonian (\ref{HK})
starts at the minimal value of $H_0(k,K)$ for the Pauli-allowed
space. This threshold value  is equal to $2\vep_F$. The
diagonalization of this Hamiltonian matrix in ${\cal H}_Q$ subspace
results either in a set of one-channel pseudostates $|z^l_k\rangle$
with energies $E_k^l$ or in a set of coupled-channel pseudostates
 $|z_k^{\vak}\rangle$ with eigenvalues $E_k^{\vak}$ both expanded over the free WP basis
 (similarly to the case of the
ordinary $NN$ scattering in a free space discussed in Section II).
Also the effective Hamiltonian may have bound-states, i.e. the
states $|z_n^b\rangle$ with energies $E_n$ which are located below
the threshold (see below).

 Thus, in the above discrete
WP-representation, the total resolvent $\hat{G}(K,W)$ from
eq.~(\ref{gkz}) can be approximated by the following superposition:
\begin{equation}
\hat{G}^{js}(K,W)\approx\sum_{n=1}^{N_b}\frac{|z_n^b\rangle\langle
z_n^b|}{W-E_n} +\sum_{\vak=1}^d\sum_{k=1}^{N_{\rm
eff}}{|z_k^\vak\rangle g_k^\vak(W)\langle z_k^\vak|},
\label{g_B_disc}
\end{equation}
where the multiplicity of the continuum $d$ is equal to 2 or 1.
$N_{\rm eff}$ is a number of pseudostates of the effective
Hamiltonian in each channel.

Finally, by using eq.~(\ref{GGK}), one gets a simple relation for
the Brueckner reaction matrix which is conveniently represented as
sum of three terms
\begin{equation}
\hat{T}=\hat{V}+\hat{T}^b+\hat{T}^c, \label{diag_g}
\end{equation}
where ${\hat T}^b$ and ${\hat T}^c$  correspond to bound-state and
continuum contributions respectively.
 In the wave-packet partial wave representation they have the
following forms:
$$
T^{b,js}_{il,i'l'}(K,W)=\sum_{n=1}^{N_b}\frac{\tilde{V}^{js}_{il,n}
\tilde{V}^{js}_{i'l',n}}{W-E_n},\eqno{(\ref{diag_g}a)}
$$
$$ 
T^{c,js}_{il,i'l'}(K,W)=
  \sum_{\vak,k}
{\tilde{V}^{js}_{il,k,\vak}
g_k^\vak(W)\tilde{V}^{js}_{i'l',k\vak}},
\eqno{(\ref{diag_g}b)}
$$
where the matrix elements of the interaction are calculated by using
the following relations
\begin{equation}
\label{vq12} \tilde{V}^{js}_{il,k\vak}\equiv \langle x_i^l|{\hat
V}{\hat Q}^{\half}|z_k^\vak\rangle= \sum_{l',i}
{V^{js}_{il,i'l'}\sqrt{Q_{i'}}C_{i'k}^{l'\vak}},
\end{equation}
and the similar formulas are for the bound-state part.

Let us mention that  it is not
necessary to employ the WP basis states in ${\cal H}_\Gamma$
subspace. The reaction matrix is determined from the eq.~(\ref{diag_g}) in momentum
representation  in which the
finite-dimensional approximation (\ref{g_B_disc}) via pseudostates
for $\hat{G}(K,W)$ is employed .

As is noted above (in particular for the so-called continuous
choice of the energy spectrum of particle and hole states) the
spectrum of the effective Hamiltonian (\ref{eqeffham}) may exhibit
eigenvalues $E_n$, which are below the threshold of the continuum:
\begin{equation}
\label{boundstates}
E_n - 2 \varepsilon_F < 0\,.
\end{equation}
Such bound states  can lead to numerical instabilities in
conventional methods for solving the BHF equations, while they just
receive special attention  in the calculational scheme presented
here according to $\hat{T}^b$ term. These bound states
 embedded in the medium of nuclear matter are also of
interest from physical point of view. They have been recently
studied in the literature (see e.g.~\cite{refboundst}). For the
calculational scheme presented here, these bound states are a kind
of spin-off product  and we will discuss them below.

\section{Some illustrative examples for symmetric nuclear matter}
The examples which we are presenting in this section have been
evaluated for isospin symmetric nuclear matter at various densities,
which are described in terms of the corresponding Fermi momentum
$k_F$. The angle-average Pauli operator (\ref{angavq}) is  employed
and we consider for this explorative presentation of the method the
$NN$ propagators with simple averages of the corresponding
two-nucleon energies, which are independent on the angle between
relative and c.m. momenta, so that the solution of the
Bethe--Goldstone equation can be done in the partial wave basis. All
results have been obtained for the CD~Bonn potential \cite{CDBonn}
considering just the proton-neutron interaction in all channels.

\subsection{Bound two-particle states}
At  first,  we want to make a few remarks on the appearance of
two-nucleon bound state configurations in the medium of nuclear
matter. As already noted above these states emerge as a kind of
byproduct in our calculations. From experiment one knows that in the
vacuum there is only one bound state, the deuteron. It is of course
a common feature of all realistic NN interactions that they
reproduce this bound state at -2.224 MeV in the $^3S_1 - ^3D_1$
channel and do not generate bound states in any other partial wave.

\begin{figure}[h!]
\centering \epsfig{file=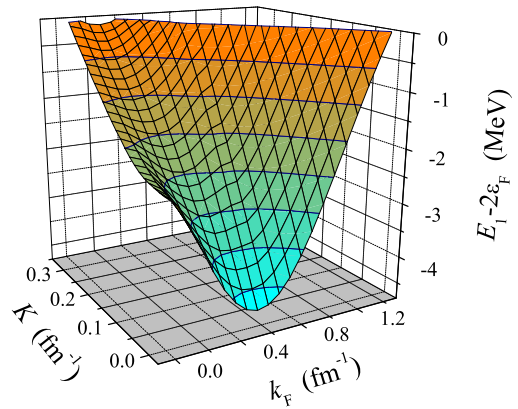,width=\columnwidth} \caption{
(Color online) The energy of the two-nucleon bound states in nuclear
matter in the deuteron channel as a function of c.m. momentum $K$
and Fermi momentum $k_F$.\label{fig1new}}
\end{figure}

Naively, one may expect that the Pauli principle reduces the
available phase space and therefore will make it more difficult to
generate bound states. However, beside the NN interaction, a very
important ingredient for the formation of bound states is the
density of states close to the threshold. This density increases in
nuclear matter with increasing Fermi momentum. Therefore the energy
of the bound state, $ E_k^{\vak} - 2 \varepsilon_F$ becomes more
attractive at low densities and exhibits a minimum of around
-4.5~MeV at $k_F$ = 0.6 fm$^{-1}$ as can be seen from the results
displayed in Fig.\ref{fig1new}. Due to the Pauli principle effect
discussed above, the binding energy gets less attractive for larger
values of $k_F$. The quasi-deuteron states, however, remains up to
$k_F$ = 1.3 fm$^{-1}$, which is just below the empirical value of
the saturation density.

The binding energy of the quasi-deuterons in the medium is very
sensitive to the c.m. momentum under consideration and decreases
very rapidly with increasing c.m. momentum $K$.

\begin{figure}[h!]
\centering \epsfig{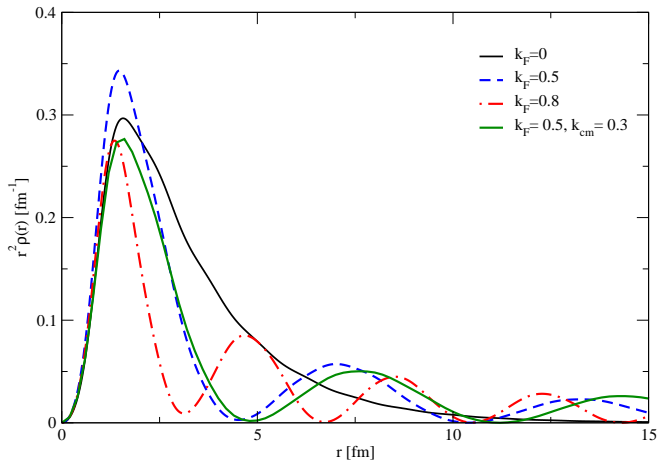} \caption{
(Color online) The density of the quasi-deuteron as a function of
relative distance $r$ for various Fermi momenta $k_F$. The results
for $k_F$ = 0.5 fm$^{-1}$ (blue dashed line) and 0.8 fm$^{-1}$ (red
dashed dotted line) have been determined for a c.m. momentum $K=0$
and are compared to the corresponding density profile of the
deuteron in free space (black solid line). For $k_F$  =  0.5
fm$^{-1}$, a profile for c.m. momentum $K$ =0.3 fm${^{-1}}$  (green
line) is presented also. Note that the density $\rho(r)$ is
multiplied by $r^2$ to enhance the density at large
$r$.\label{fig2new}}
\end{figure}

Examples for the density profile for quasi-deuteron states are
displayed in Fig. \ref{fig2new} as  functions of the distance
between the nucleons and compared to the corresponding density
distribution for the ``free''\ deuteron. The most pronounced
difference between the density profiles for the deuteron and the
bound-state structures in the nuclear medium are the ripples in the
latter. The scale of these ripples is determined by the relative
momenta slightly above the Fermi momentum, which dominates the
momentum distribution in the wave-functions of these states. The
decrease of the density with increasing $r$, on the other hand,
reflects the binding energy of the states and it  is weaker for the
states with the larger c.m. momentum.

All these bound-state structures in the medium discussed so far have
been evaluated  assuming a pure kinetic energy for the
non-interacting nucleons also in the nuclear medium. Inclusion of
the single-particle self-energy tends to reduce the binding energies
of the bound-state structures, as the self-energy lowers the density
of states around the Fermi energy. The momentum dependence of the
self-energy is often expressed in terms of an effective mass $m^*$
(see the eq.~(\ref{efmas}) and discussion below). A detailed
evaluation of the single-particle potential yields an enhancement of
this effective mass close to the Fermi energy ($m^*\to m$), which
originates mainly from the energy-dependence of the self-energy
($E$-mass) \cite{Muether_02} (see also discussion below). Therefore
the calculations ignoring the effects of the single-particle
potential may  be not too bad, since the effects of the
single-particle potential drops out when they are in the difference
$E_B=E_n - 2 \varepsilon_F$. In fact, while the calculation of the
quasi-deuteron in the nuclear medium at $k_F$ = 1.0 fm$^{-1}$
($K=0$) yields an energy of -2.9 MeV when calculated without
self-energies included, the corresponding calculation considering
the complete momentum- and energy-dependence of the self-energy
yields -1.8 MeV.

\begin{figure}[h!]
\centering \epsfig{file=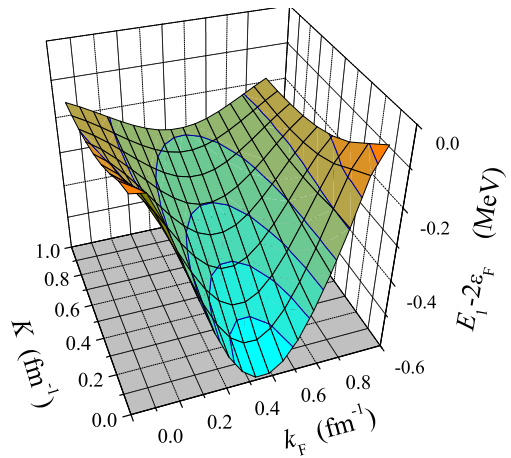,width=\columnwidth} \caption{The
energy of the bound two-neutron states in $^1S_0$ channel as a
function of c.m. momentum and the Fermi momentum.\label{fig3new}}
\end{figure}

While there is only one bound state in the vacuum, the deuteron, in
the nuclear medium one may also obtain a bound state in other
partial waves. In our calculations, we observe bound states also in
the $^1S_0$ channel with isospin $\tau=1$. From the results
displayed in Fig.~\ref{fig3new} one can see that the CD~Bonn
interaction yields bound-state configurations in this channel for
Fermi momenta $k_F$ up to 1 fm$^{-1}$ with a maximal binding energy
of 0.6 MeV, which is considerably weaker than in the case of the
quasi-deuteron.

It should be mentioned that the occurrence of quasi-bound states
discussed here typically requires more attractive residual
interaction than a corresponding ``pairing instability'', which has
been discussed for the evaluation of the nuclear $T$-matrix within
the theory of self-consistent Greens function \cite{frick}, as the
evaluation of this $T$-matrix also encounters the contribution of
hole-hole ladders.

\subsection{Discussion of the single-particle self-energy}
One of the major advantages of the present scheme for solving the
Bethe--Goldstone equation (\ref{BG}) is its efficiency whenever the
 reaction matrix is needed for various values of the energy parameter $W$.
Therefore it is very easy with this approach to evaluate the whole
energy- and momentum-dependence of the real and imaginary part of
the single-particle potential or nucleon self-energy $U(k_1,\omega
)$ as defined in (\ref{ukk}).

The explicit formula for the self-energy (\ref{ukk}) in
 a case of the angle-averaged Pauli-projector can be
rewritten in a form:
\begin{eqnarray}
 U(k_1,\omega)=\frac{2}{k_1}\sum_{jsl}(2j+1)(2\tau+1)
\int_{\Om(k_1)} dk dK kK\nonumber\\
T^{js}_{ll}\left(k,k;K,W=[\omega+H_0(k,K)-e(k_1)]\right),\label{u_rel}
\end{eqnarray}
where the integration domain $\Om(k_1)$ over $k$ and $K$ depends on
the sp momentum $k_1$ value \cite{Tabakin}. By using a discrete wave-packet
representation for the reaction matrix and also some integration mesh $\{K_n\}$ with
 weights $\{\De K_n\}$ for the c.m. momentum, the explicit relation takes
the form of a discrete sum:
\begin{eqnarray}
U(k_1,\tilde{\omega}+e(k_1))=\frac{2}{k_1}\sum_{jsl}(2j+1)(2\tau+1) \nonumber\\
\sum_{i,n}T^{js}_{il,il}(K_n,\tilde{\omega}+H_0(k_i,K_n) )\De K_n.
\end{eqnarray}
Here matrix elements $T^{js}_{il,il}$ are defined by the
eqs.~(\ref{diag_g}). It is straightforward to extract explicitly the
Hartree--Fock part (caused by the bare interaction $\hat V$ only) and the
bound-state part from the self-energy by using this explicit
formula. We note also that the diagonalization procedure should be
done for every value of $K_n$ and the matrix elements for different
$k_i$ and $\tilde{\omega}$ are calculated by using just the same
system of pseudostates of the effective Hamiltonain.

\begin{figure}[h]
\centering \epsfig{file=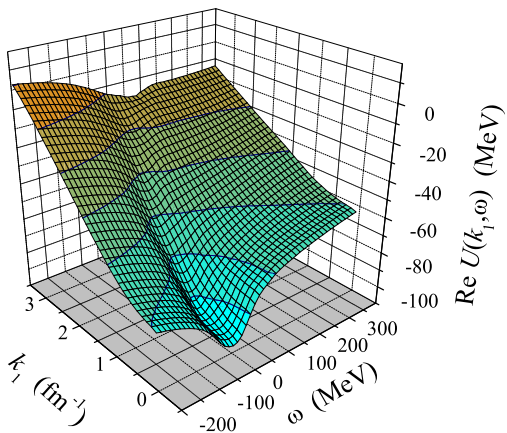,width=\columnwidth} \caption{
(Color online) Real part of nucleon self-energy as a function of
momentum and energy, calculated for a Fermi momentum $k_F$ = 1.3
fm$^{-1}$. \label{fig:real}}
\end{figure}

\begin{figure}[h]
\centering \epsfig{file=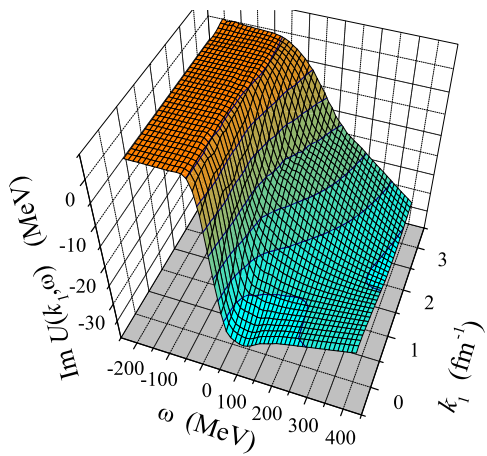,width=\columnwidth} \caption{
(Color online) Imaginary part of nucleon self-energy as a function
of momentum and energy, calculated for a Fermi momentum $k_F$ = 1.3
fm$^{-1}$.\label{fig:imag}}
\end{figure}

As an example we show in Figs.~\ref{fig:real} an \ref{fig:imag} the
real and the imaginary parts of the self-energy $U$ calculated at
$k_F$ = 1.3 fm$^{-1}$ for various sp momenta $k_1$ and energies
$\omega$. The pole structure in the Bethe--Goldstone equation leads
to an imaginary part only at energies $\omega$ above the Fermi
energy $\varepsilon_F$, which is around -37 MeV in our example. This
pole structure is also important for the energy dependence of the
real part of $U$. It  is seen that ${\rm Re}\  U(k_1,\omega)$ is
decreasing when the energy $\omega$ is rising being  negative, has a
minimum around $\omega = 0$ and then it rises at positive values of
$\omega$.

\begin{figure}[h]
\centering \epsfig{file=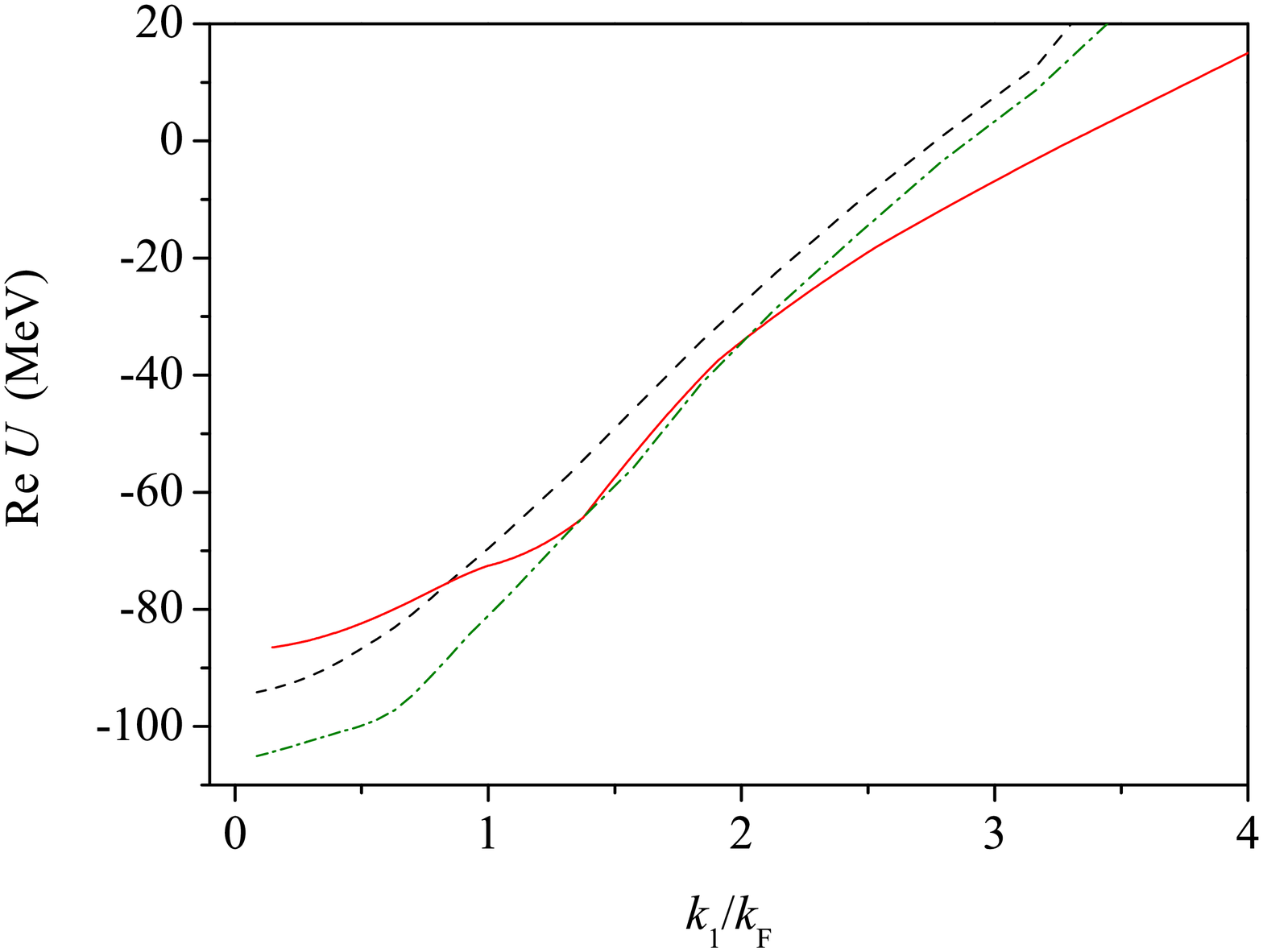,width=\columnwidth} \caption{Real
part of the self-energy, calculated at a fixed energies $\omega=-50$
MeV (dashed curve) and $\omega=0$ MeV (dash-dotted curve)  and found
self-consistently following eq.~(\protect{\ref{ukk}}) (solid curve).
Example for a Fermi momentum $k_F$ =1.3 fm$^{-1}$ \label{fig:uofk}}
\end{figure}

As a consequence, the momentum dependence of the single-particle
potential $U(k_1, \omega)$ at a fixed value of $\omega$ is much
stronger than that for the self-consistent definition of the energy
variable according to eq.~(\ref{ukk}) if we consider momenta below
or slightly above the Fermi momentum. This is visualized in
Fig.~\ref{fig:uofk}, in which we compare results for ${\rm Re}\
U(k_1,\omega)$ for different energies $\omega$ with a
single-particle potential found self-consistently.

This feature is well known and has been discussed in the literature
as an enhancement of the effective mass at the Fermi energy due the
energy-dependence of the self-energy or the so-called $E$-mass
effect~\cite{Muether_02,emass2}

\subsection{Calculation of the sp potential and the equation of state}
As another example, we calculate the self-consistent sp potential
and the equation of state by using the simple effective mass
approximation for the sp energy at each iteration step. However, to
check the reliability of this technique, two types of such an
approximation have been employed, which differ by the maximal
single-particle momentum value $k_{1}^{\rm max}$ used in the fitting
of the effective mass parameters: $k_{1}^{\rm max}=1.5k_F$ (the
calculation 1) and  $k_{1}^{\rm max}=4k_F$ (the calculation 2).

\begin{figure}[h]
\centering \epsfig{file=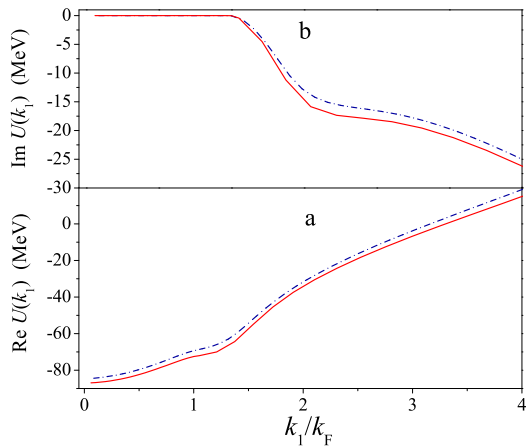,width=\columnwidth}
 \caption{ (Color online) Real (a) and
imaginary (b)  parts of the sp potential calculated
self-consistently by using the effective mass approximations
according to the calculation 1 (dash-dotted curve) and 2 (solid
curve).  \label{fig:ab}}
\end{figure}

It turns out that the case 1 in which one makes use the smaller
fitting interval leads to a slightly smaller effective mass ($m^*/m$
= 0.694 at $k_F$ = 1.3 fm$^{-1}$) than in the case 2 where one gets
 $m^*/m$ = 0.746 (here $m$ is the nucleon mass).
 In Figs.~\ref{fig:ab}a and b the real and imaginary parts of the
sp potentials calculated from self-consistent iterations by using
these two approximations in energy terms $H_0$ are represented. One
observes clearly that the shape in these figures is rather similar.
The calculation 2, however, yields slightly larger absolute values
for the real and imaginary part. The differences seem to be small in
this figure. Note, however, that the real part of the self-energy in
calculation 2 is up to 2 MeV more attractive than the corresponding
values for calculation 1. 

\begin{figure}[h]
\centering \epsfig{file=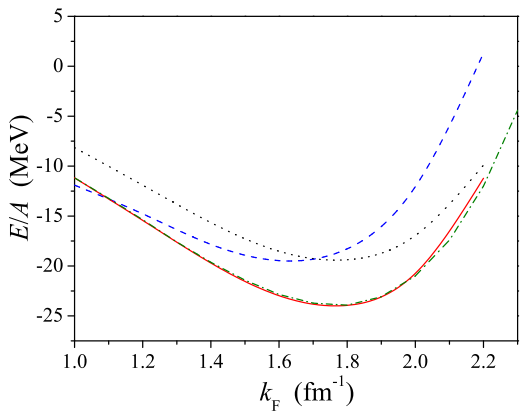,width=\columnwidth} \caption{
(Color online) Binding energy per nucleon calculated    via the
wave-packet diagonalization  technique for the effective mass
approximations in calculations 1 (dashed curve) and 2 (solid curve)
in comparison with the results for the conventional sp spectrum
choice (dotted curve). The dash-dotted curve corresponds to the
results of the fully self-consistent calculations for the
continuous choice from the refs.~\cite{Muether_02,Gad}.
    \label{fig:e/a}}
\end{figure}

These differences become even more visible in the  binding energy dependence of
the Fermi momentum $k_F$  obtained from the corresponding sp potentials by using
the formula (\ref{eq.EA}). In Fig.~\ref{fig:e/a}, we compare the
results of our discrete wave-packet technique using  the above two
approximations 1 and 2  with the results of the fully self-consistent
approach from the refs.~\cite{Muether_02,Gad}, where
angle-independent energy terms in the BGE with the exact
non-averaged Pauli projector have been used. The agreement of the
latter results with the EOS found in the WP approach  for the
calculation 2 is rather well. While the calculation 1 with a shorter fitting interval for the effective mass approximation  results in a smaller binding energy. This reflects a real problem with the
effective mass approximation.

Thus, we have demonstrated clearly that the diagonalization
technique developed here is very useful  for  evaluation of the
Brueckner reaction matrix and the single-particle spectrum in
nuclear matter at various densities. However the accurate treatment
of the nuclear matter EOS still requires also calculation of three-
and few-body correlation contributions for the binding energy in
nuclear matter. This leads to solving complicated three-body
equations for the reaction matrix such as the Bethe--Faddeev ones. A
direct solution of the latter equations for realistic $NN$- and
$3N$-interactions in fully self-consistent scheme is a very hard
task nowadays. However, it seems that some effective three-body
Hamiltonian  can be defined for the three-body system in the
Pauli-allowed subspace (in much the same way as  the two-body one
from the eq.~(\ref{HK})), so that the diagonalization technique
might be generalized for a proper account of three-body correlations
as well. Such an approach will simplify enormously the evaluation of
 three-body force contribution in a traditional scheme.
\section{Summary}
In the present work, we have demonstrated that  the accurate
multienergy solution for the Lippmann--Schwinger integral equation
for single- and coupled-channel $t$-matrix can be easily found from
the direct single diagonalization procedure for the total
Hamiltonian matrix in $L_2$ basis of the stationary wave packets.
This approach has been tested carefully for two particular $NN$
interaction models both for a single-channel and also
coupled-channel transition operators. In all the cases, a very good
accuracy of the direct diagonalization procedure as compared to
 solution of the respective integral equation has been
attained. Thus, this approach provides a very  efficient way for
finding the  coupled-channel off-shell $t$-matrix at various
energies.

This important feature is especially valuable for solving the
few-body scattering problems where the kernel of the Faddeev-like
equation includes fully off-shell $t$-matrix at many energies.

 At
the next step, we have generalized this approach to a solution of
the Bethe--Goldstone integral equation and derived an explicit form
of the effective Hamiltonian in the Pauli-allowed two-particle
subspace. Thus, the multiple numerical solutions for the
Bethe--Goldstone integral equation for the reaction matrix at
different values of the relative momentum and energy  have been
replaced by a single matrix diagonalization of the effective
Hamiltonian in the Pauli-allowed subspace which is much simpler and
faster.

The method can be extended to  modern  modifications of the BHF
approach which include more complicated forms of the particle and
hole propagators (such as the $pphh$ propagator \cite{book}),
non-zero temperature regime etc. Moreover, this direct
diagonalization technique might open a door to realiable and
accurate treatment of three- and few-body correlations in dense
nuclear matter.

{\bf Acknowledgment:} The authors appreciate the partial financial
support of the DFG grant MU 705/10-1, the joint DFG--RFBR grant
16-52-12005 and the RFBR grant 16-02-00049.

\appendix
\section{Eigenvalues of the coupled-channel total resolvent  in the WPCD approach}
\subsection{Approximation for the total resolvent in a single-channel case}
To find the eigenvalues for the total resolvent in the pseudostate
basis, let us remind some results from the general WPCD approach
\cite{Annals}.

The scattering wave packets  for some Hamiltonian $\hat h$ are
constructed as integrals of the exact scattering wave-functions
$|\psi(E)\rangle$ over some discretization intervals
$[\vep_{i-1},\vep_i]_{i=1}^M$, (similarly to free WPs):
\begin{equation}
\label{zi}
|\bar{z}_i\rangle=\frac{1}{\sqrt{\Delta_i}}\int_{\vep_{i-1}}^{\vep_i}dE|\psi(E)\rangle,\quad
\Delta_i=\vep_i-\vep_{i-1}.
\end{equation}
These scattering WP states, jointly with the possible boundstates
$|z_n^b\rangle$ of the Hamiltonian, form a WP-space for the
Hamiltonian $\hat{h}$
 with the projector \cite{Annals}:
\begin{equation}
\mathfrak{p}=\sum_{n=1}^{N_b}|z^b_n\rangle\langle
z^b_n|+\sum_{i=1}^M|\bar{z}_i\rangle\langle \bar{z}_{i}|,
\end{equation}
where $N_b$ is a number of bound states. The Hamiltonian can be
approximated as a finite sum in such a WP-space:
\begin{equation}
\label{php} \hat{h}\approx
\mathfrak{p}\hat{h}\mathfrak{p}=\sum_{n=1}^{N_b}|z^b_n\rangle\vep^b_n\langle
z^b_n|+\sum_{i=1}^M|\bar{z}_i\rangle\bar{\vep}_i\langle\bar{z}_i|,\quad
\end{equation}
where $\vep^b_n$  and the midpoints
$\bar{\vep}_i=\half[\vep_{i-1}+\vep_i]$ represent eigenvalues of the
total Hamiltonian for its bound states and discretized continuum
states correspondingly. Then, the finite-dimensional approximation
for the total resolvent in the basis built takes the same diagonal
form:
\begin{equation}
\label{pgp} \hat{g}(E)\approx
\mathfrak{p}\hat{g}(E)\mathfrak{p}=\sum_{n=1}^{N_b}\frac{|z^b_n\rangle\langle
z^b_n|}{E-\vep^b_n}+\sum_{i=1}^M|\bar{z}_i\rangle g_i(E) \langle
\bar{z}_i|,\end{equation}
 where eigenvalues $g_i(E)$ are expressed as follows \cite{Annals}:
\begin{equation}
\label{gie}
g_i(E)=\frac1\Delta_i\left[\ln\left|\frac{E-\vep_{i-1}}{E-\vep_i}\right|-{\rm
i}\pi\theta\left(E\in[\vep_{i-1},\vep_i]\right)\right].
\end{equation}
Here the generalized Heaviside-type theta-function is introduced:
\begin{equation}
\theta\left(E\in[\vep_{i-1},\vep_i]\right)=\left\{
\begin{array}{cc}
1,&E\in[\vep_{i-1},\vep_i],\\
0,&E\notin[\vep_{i-1},\vep_i].\\
\end{array}\right.
\end{equation}
It should be stressed, that the formula (\ref{gie}) is universal for
any Hamiltonian for which the WP states can be constructed. Thus, it
is also valid for a free resolvent $g_0(E)$ eigenvalues in the free
WP basis (\ref{xi}).

The diagonalization procedure for the total Hamiltonian matrix $h$
in free WP basis $\{|x_i\rangle\}_{i=1}^N$ results in a finite set
of eigenvectors  $\{|z_i\rangle\}_{i=1}^N$ with the respective
eigenenergies $\{E_i\}$. Assume further that these eigenfunctions
are enumerated in order of increasing  the eigenvalues. If there is
a bound state in the system, the first state $|z_1\rangle$ with
negative energy $E_1$ is assumed to be an approximation for this
bound state wave function $|z_b\rangle$, while all the other
eigenfunctions with positive eiegenvalues are pseudostates
representing somehow the total Hamiltonian continuum. It has been
shown previously \cite{Annals} that these normalized pseudostates
can be considered as approximations for scattering WPs (rather than
approximations for non-normalized
 scattering wave functions).
 So that, one can
replace  exact scattering WP functions in the eq.~(\ref{pgp}) with
corresponding pseudostates. Finally, the  pseudostate-approximation
(\ref{g_pseudo})
 for the total resolvent is found
in which the exact eigenvalues (\ref{gie}) for scattering WPs are used.

The only problem which arises here is how to construct the
discretization mesh $\{\vep_i\}$ in such a way that pseudostate
energies $\{E_i\}$  are coincided with eigenvalues $\bar{\vep}_i$
from eq.~(\ref{php})  in the exact scattering WP basis. Such a
reconstruction of the intervals $[\vep_{i-1},\vep_i]_{i=2}^N$ can be
done approximately by the following way:
\begin{eqnarray}
\label{vep} \vep_0=0,\quad \vep_i=\half[E_i+E_{i+1}],\quad
i=1,\ldots,N-1,\quad
\nonumber\\
\vep_N=E_N+\half(\vep_{N-1}-\vep_{N-2}).\qquad \qquad
\end{eqnarray}
Here the midpoints $\bar{\vep}_i$ of the reconstructed bins differ a
little bit from the exact $E_i$ values, however, with increasing the
basis dimension, this difference becomes smaller and does not cause
visible errors in the whole solution.

Thus, the eigenvalues of the total resolvent in the pseudostate
basis can be found by using the formula (\ref{gie}) in which
endpoints of the energy intervals are calculated from pseudoenergies
$E_i$ using the  eq.~(\ref{vep}).

\subsection{Coupled-channel pseudostates and the total resolvent eigenvalues}
In case of  coupled-channel total Hamiltonian $\hat h$, the
scattering wave packets are constructed from the exact scattering
wave functions $|\psi^\vak(E)\rangle$ defined in the eigenchannel
representation. So, for this purpose, the continuous spectra in
eigen channels $\vak$=1 and 2 are divided onto intervals
$\{[\vep_{i-1}^\vak,\vep_{i}^\vak]_{i=1}^{M^\vak}\}_{\vak=1}^2$ and
the coupled-channel scattering WPs are introduced:
\begin{equation}
\label{z_vak}
|\bar{z}_i^\vak\rangle=\frac1{\sqrt{\Delta_i^\vak}}\int_{\vep_{i-1}^\vak}^{\vep_i^\vak}dE
|\psi^\vak(E)\rangle,\quad
\Delta_i^\vak=\vep_i^\vak-\vep_{i-1}^\vak,
\end{equation}
similarly to the one-channel case.

Further, one adds the possible bound states and introduces the
WP-space for the total coupled-channel Hamiltonian, similarly to
one-channel case. At last, one gets the following WP-approximation
for the total coupled-channel resolvent
\begin{equation}
\label{pgp_m} \hat{g}(E)\approx
\sum_{n=1}^{N_b}\frac{|z_n^b\rangle\langle
z^b_n|}{E-E_n}+\sum_{\vak=1}^2\sum_{i=1}^{M^\vak}|\bar{z}_i^\vak\rangle
g_i^\vak(E) \langle \bar{z}_i^\vak|,\end{equation} where eigenvalues
$g_i^\vak(E)$ are defined by the formula (\ref{gie}) in which the
$\vak$-channel interval bound-points $\vep_i^\vak$ should be used.

To treat accurately coupled-channel pseudostates, we have shown
previously \cite{Annals,KPRF} that the free WP basis states in the
initial unperturbed channels (e.g. channels corresponding to the
fixed orbital momentum $l$ value) should be constructed in such a
way that the free Hamiltonian matrix has multiple discrete
eigenvalues \cite{Annals,KPRF}. In such a case, the spectrum of the
total Hamiltonian matrix $\bf h$  consists of pairs of slightly
shifted nearby eigenenergies (except possible bound states). Thus,
this spectrum can be separated onto two branches. Finally, these two
separated branches of eigenvalues  are considered as discretized
eigenchannel spectra.

For the two-channel case discussed in the present paper, this
separation is done just by dividing the eigenvalues of the total
Hamiltonian matrix with even and odd indices. Further, the
discretization end-points are  built for each of two eigenchannels
$\vak=1,2$ separately from the eigenvalues $E_i^\vak$ by using the
algorithm similar to the single-channel one (\ref{vep}).


\begin{thebibliography}{99}
\bibitem{newton} R.G. Newton, {\em Scattering Theory of Waves and
Particles}
(McGraw-Hill, New York, 1966).
\bibitem{Tabakin}M. I. Haftel and F. Tabakin, Nucl. Phys. A {\bf 158}, 1 (1970).
\bibitem{Muether99}E. Schiller, H. M{\"u}ther, P. Czerski, Phys. Rev. C {\bf 59}, 2934
(1999).
\bibitem{Baldo} M. Baldo et al., Phys. Rev. C {\bf 65}, 017303
(2001).
\bibitem{book} W.H. Dickhoff, D. van Neck, {\em Many-Body Theory Exposed!
Propagator description of quantum mechanics in many-body systems}
(World Scientific, 2005).
\bibitem{Muether_02} T. Frick, Kh. Gad,  H. M{\"u}ther, P. Czerski,
Phys. Rev. C {\bf 65}, 034321 (2002).
\bibitem{Sartor} R. Sartor, Phys. Rev. C {\bf 73}, 034307 (2006).
\bibitem{Annals} O.A. Rubtsova, V.I. Kukulin, V.N. Pomerantsev, Ann.
Phys. {\bf 360}, 613 (2015).
\bibitem{Heller} E.J. Heller, T.N. Rescigno, W.P. Reinhardt, Phys. Rev. A {\bf 8}, 2946 (1973).
\bibitem{reinhardt}J.R. Winick, W.P. Reinhardt, Phys. Rev. A {\bf 18}, 910 (1978),
{\it ibid.} {\bf 18}, 925 (1978).
\bibitem{papp} Z. Papp, C-.Y. Hu, Z.T. Hlousek, B. K{\'o}nya,  S.L.
Yakovlev, Phys. Rev. A {\bf 63} (2001) 062721.
\bibitem{LIT} S. Quaglioni, W. Leidemann, G. Orlandini, N. Barnea,
V.D. Efros, Phys. Rev. C {\bf 69} (2004) 044002.
\bibitem{dingri2016} D. Ding, A. Rios, H. Dussan, W.H. Dickhoff, S.J. Witte, A.Polls, arXiv:1601.01600.
\bibitem{KPRF} O.A. Rubtsova, V.I. Kukulin, V.N. Pomerantsev, A.~Faessler, Phys. Rev. C {\bf 81},
064003 (2010).
\bibitem{validity} O.A. Rubtsova, V.I. Kukulin, V.N. Pomerantsev, Phys. Rev.
 C {\bf 84}, 044002 (2011).
\bibitem{Greiner} M. Danos, W. Greiner, Phys. Rev. {\bf 146}, 708 (1966).

\bibitem{CDBonn} R. Machleidt, F. Sammarruca, and Y. Song, Phys. Rev. C {\bf 53}, R1483
(1996).
\bibitem{Nijm} V.G.J. Stoks, R.A.M. Klomp, C.P.F. Terheggen,
J.J~de~Swart, Phys. Rev. C {\bf 49}, 2950  (1994).
\bibitem{BBP} H.A. Bethe, B.H. Brandow, and A.G. Petchek, Phys. Rev. {\bf 129}, 225 (1963).
\bibitem{3-body} H.Q. Song, M. Baldo, G. Giansiracusa, and U. Lombardo,
Phys. Rev Lett. {\bf 81}, 1584 (1998).
\bibitem{mahaux} J.P. Jeukenne, A. Lejeune, and C. Mahaux, Phys. Rev.  C {\bf 16}, 80 (1977).
\bibitem{refboundst} F. Isaule, H.F. Arellano, and A. Rios, arXiv:1602.05234.
\bibitem{frick} T. Frick and H. M\"uther, Phys. Rev. C {\bf 68}, 034310 (2003).
\bibitem{emass2} C. Mahaux and R. Sartor, Adv. Nucl. Phys. {\bf 20}, 1 (1991).
\bibitem{Gad} Kh. Gad, Nucl. Phys. A {\bf 747}, 655 (2005).
\end{thebibliography}
\end{document}